\documentclass[]{interact}
\usepackage[parfill]{parskip}

\usepackage{epstopdf}
\usepackage[caption=false]{subfig}

\usepackage[authoryear]{natbib}
\usepackage{url}

\theoremstyle{plain}

\theoremstyle{definition}

\theoremstyle{remark}

\usepackage{xcolor}

\usepackage{multicol,ragged2e}
\usepackage{threeparttable}
\usepackage{siunitx} 

\begin{document}

\title{De-meaning Simulation Studies}

\author{
\name{Thomas Lumley\textsuperscript{a} and Brian Williamson\textsuperscript{b} and Pamela Shaw\textsuperscript{b} }
\affil{\textsuperscript{a}Department of Statistics, University of Auckland, Auckland, New Zealand; \textsuperscript{b}Biostatistics Division, Kaiser Permanente Washington Health Research Institute, Seattle, Washington}
}

\maketitle

\begin{abstract}
In simulation studies evaluating asymptotic approximations it is common practice to report averages and standard deviations over repeated simulations.  We argue that quantile-based summaries are more appropriate from both a theoretical and practical point of view.   Theoretically, convergence of moments --- or even existence of moments --- is not guaranteed by convergence in distribution, so sample moments are not ideal for assessing the accuracy of a distributional approximation.  In practice, means and variances are not good summaries of approximately-Normal distributions that may have occasional outliers.  We suggest the median and median absolute deviation, and empirical confidence interval coverage, as better general summaries, and argue that moments should be reserved for simulation settings where they are of substantive interest. 
\end{abstract}


\section{Introduction}

It is common when developing statistical methods to conduct simulation studies and report, for an estimator $\hat\theta$ of a parameter $\theta$, the mean and variance (or standard deviation) of $\hat\theta$ over a large number of simulations  \citep[for example:][]{sim-with-means-1,sim-with-means-2,sim-with-means-3,sim-with-means-4,sim-with-means-5,sim-with-means-6,sim-with-means-7}.  These moment estimators are used to evaluate the quality of estimators. For example, $\hat\theta$ would be compared to the true value of $\theta$ for that simulation study and the mean of the standard error (SE) estimates would be compared to the empirical SE of $\hat\theta$ across the simulations. 

In this paper we argue that quantile-based simulation summaries should often be preferred.  First, as a matter of  data analysis practice we should not default to reporting moments, because they are sensitive to arbitrarily rare outliers (that is, they have a \emph{breakdown point} of zero and high \emph{gross-error sensitivity} \citep{huber-book, hampel-book}). Second, when the simulations are motivated by asymptotic arguments for convergence in distribution we should prefer summaries whose behaviour allows us to assess convergence in distribution.

In one very common setting, we have theory showing asymptotic normality of an estimator $\hat\theta_n$, ie,
$$\sqrt{n}(\hat\theta_n-\theta_0)\stackrel{d}{\to} N(0,\sigma^2).$$
We are interested in how close the finite-sample distribution of $\hat\theta_n$ is to $N(\theta_0,\sigma^2/n)$. Given a variance estimator $\hat\sigma^2$, we are also interested in how close $\hat\sigma^2$ is to $\sigma^2/n$, and whether an interval estimator $[\hat\theta_n-1.96\hat\sigma,\hat\theta_n+1.96\hat\sigma]$ has close to its nominal coverage. 

If simulations output realizations of $\hat\theta^*_m\sim N(\mu,\sigma^2/n)$,  the sample median is a consistent (and unbiased) estimator of $\mu$, and the rescaled MAD
$$\mathrm{MAD}(X) = \mathrm{median}\{ |X-\mathrm{median}{X}|\}/\Phi^{-1}(3/4)$$
is an estimator of $\sigma/\sqrt{n}$ such that $\mathrm{MAD}^2$ is unbiased for $\sigma^2/n$ and $n\mathrm{MAD}^2$ is consistent for $\sigma^2$.   

If we weaken the assumption to asymptotic normality, ie, $\sqrt{n}(\hat\theta^*_m-\mu)\stackrel{d}{\to} N(0,\sigma^2)$, the median and MAD still perform as expected but the mean $\bar \theta^*$ and the variance $S_{\hat\theta}^2$ need not. Convergence in distribution implies convergence of quantiles but does not imply convergence (or even existence) of moments \citep{proschan18}.   If the question for the simulations is``Have we reached asymptopia?'', ie, ``is $n$ large enough for the $N(\mu,\sigma^2)$ approximation to the distribution to be useful?", assessing moments is inappropriate.

\section{Examples}
Here we provide illustrations of the two problems that can result from using simulation moments to evaluate closeness in distribution, both of which are avoided by robust summaries.  In the first example, the simulation moments appear well-behaved at any reasonable number of simulation replicates but are actually undefined; in the second, the simulation moments suggest the asymptotic approximations are worse than they actually are.  The code for both simulations is available at 
\url{https://github.com/tslumley/demeaning/}

\subsection{ Logistic regression}

Suppose $X$ is a predictor variable with a standard Normal distribution and $Y$ a binary outcome, with $$\mathrm{logit}\,P[Y=1|X=x]=\alpha+\beta x.$$
We observe $n$ independent and identically distributed (iid) observations of $(X,Y)$ and estimate $\alpha$ and $\beta$ by maximum likelihood.   The limiting Normal distribution of $\sqrt{n}(\hat\alpha-\alpha,\hat\beta-\beta)$ has zero mean and diagonal covariance matrix with diagonal values $(4n^{-1},4n^{-1})$. 

Table~\ref{logistic} shows the mean and median, and the standard deviation and MAD from 10,000 simulations when $X\sim N(0,1)$ and $(\alpha,\beta)=c(0,0)$ and $n=100$.  The values for the median and MAD are correct to within about 0.002 and 0.01, respectively. The values for the mean and standard deviation, however, are seriously incorrect. 

If all the values of $x$ for observations with $y=1$ are greater than all the values of $x$ with $y=0$, the MLE is $\hat\beta=\infty$, whereas if all the values of $x$ with $y=1$ are smaller, the MLE is $\hat\beta=-\infty$. The probability of each of these in a sample of size $n$ is 
$$\sum_{r=0}^{n} P\left(\sum_i Y_i=r\right) \binom{n}{r}   ^{-1}=n\times 2^{-n}>0.$$
For $n=100$ this is $7.8\times 10^{-29};$
 the distribution of $\hat\beta$ \emph{has no finite moments.}  

 If we were seriously interested in estimating $E[\hat\beta]$ or its variance, the fact that these quantities do not exist would be important, and we would need to do a large enough simulation study to demonstrate that they did not exist.   In practice we do not usually care that $\hat\beta$ can be badly behaved with very small probability. It does not matter whether $\hat\beta$ has  finite moments in finite samples as long as its distribution can be approximated well by a Normal distribution. 
 
\begin{table}
\caption{Simulation estimates of location and scale for logistic regression coefficient based on 100,000 replicates}
\label{logistic}
\centering\begin{tabular}{rr}
\hline
 \multicolumn{2}{c}{Location ($\times 1000$)}\\
mean & median \\
-0.34 & -0.22\\
\hline
\multicolumn{2}{c}{Scale}\\
 sd & MAD\\
0.21 & 0.21\\
\hline
\end{tabular}
\end{table}

 \subsection{Incomplete data models}
 

We used the synthetic-data simulation procedure described in \citet{williamson2025assessing} to further illustrate why quantile-based summaries should be preferred in a setting with incomplete data. This simulation procedure was designed to evaluate methods to handle missing data in the setting of rare outcomes.  In this setting, we have a binary outcome $Y$ with 1\% marginal outcome probability, a binary exposure of interest $X$ with 40\% marginal exposure probability, and four confounder variables ($Z_1, Z_2, W_1, W_2$). Two confounder variables ($W_1$ and $W_2$) are jointly missing with marginal probability 80\%. We generated $(X_\text{latent}, Z_1, Z_2, W_1, W_2) \sim N(0, \Sigma)$, where $X_\text{latent}$ is a latent continuous variable used to determine observed exposure $X$. The covariance matrix $\Sigma$ is defined so that all variables have variance 1, $Z_2$ and $W_2$ have correlation 0.4 with $X_\text{latent}$, and all other variables have pairwise correlation 0.2. The outcomes are generated according to 
\begin{align}\label{eq:outcome_model}
    \text{logit} P(Y = 1 \mid X = x, W = w, Z = z) =& \ -5.1 + \log(1.5)x + \log(1.5)w_1 \\
    &-\log(1.75)w_2 + \log(1.5)z_1 - \log(1.3)z_2 \notag,
\end{align}
while the missing data indicator $R$ was generated according to
\begin{align*}
    \text{logit}P(R = 1 \mid X = x, Z = z, Y = y) =& \ 1.08 + \log(2.5)x + \log(1.5)z_1 \\
    &+ \log(1.5)z_2 + \log(2.5)y.
\end{align*}
We set $W_1$ and $W_2$ to missing if $R = 1$. This missingness model is missing at random (MAR) with respect to the sampling model \citep{seaman2013review}. 

Similar to \citet{williamson2025assessing}, for this rare outcome setting, we consider the relative performance of the following estimators of the conditional log odds ratio for the exposure $X$ (i.e., the regression parameter for $x$ in Equation~\ref{eq:outcome_model}): generalized raking \citep[GR; ][]{deville1992calibration}, inverse probability weighting \citep[IPW; ][]{sarndal2003model}, multiple imputation (MI) via chained equations \citep[MICE; ][]{van2011mice}, and targeted maximum likelihood estimation \citep[TMLE; ][]{rose2011targeted}. Each of these methods is described further in \citet{williamson2025assessing}.

For each of 2500 Monte-Carlo replications, we generated a dataset of size 12,000 and 17,000 according to the above specification and fit each of the four estimators of interest. For each algorithm, we first computed performance across all simulation replicates where the algorithm converged. These include the mean bias (difference between the point estimate and the truth), empirical standard error (ESE, the standard error of the point estimates across replications), and the mean ASE. Next, we computed more robust measures of performance: the median bias and ASE; the median absolute deviation (MAD); the Winsorized mean bias and ASE (Winsorizing at the 2.5 and 97.5\% thresholds); and the trimmed mean bias and ASE (trimming at the 2.5 and 97.5\% thresholds). Based on these, we computed the ratio of median bias to median ASE. Finally, we computed two types of confidence interval coverage probabilities. The first coverage probability (CP) treats non-convergence as not covering the truth. The second (CP*) considers coverage among only datasets where all algorithms converged.


We present the results of the simulation for N=12,000 and 17,000 in Tables~\ref{tab:synthetic_results_cOR_n12000} and \ref{tab:synthetic_results_cOR_n17000}, respectively. At sample size 12,000, the GR estimation procedure failed to converge in 13\% of simulations (325 simulations) and returned an unreasonably large point estimate ($> \log 10$) in 0.7\% of simulation replicates; IPW and TMLE both never failed, but returned an unreasonable point estimate in 4.1\% and 4.75\% of replicates, respectively. At sample size 17,000, GR failed to converge in 6.8\% of simulations; GR, IPW, and TMLE returned unreasonable point estimates in 0.1\%, 1\%, and 1.8\% of simulations, respectively. In all cases, the failure of GR and the large-magnitude point estimates occurred due to the small number of observed events leading to strata of the covariates with either no events or no non-events. In this case, there was apparent bias and the mean ASE amongst the convergent datasets had poor agreement with the ESE for GR, IPW and TMLE; multiple imputation was unaffected because filling in the missing data allowed more events to be used in estimation. Based on these non-robust summaries, we may conclude that the GR, IPW and TMLE or their variance estimators are biased, particularly for the N=12,000 scenario, where the bias/SE ratio was an order of magnitude larger than $1 / \sqrt{n_{\text{eff}}}$ , where $n_{\text{eff}}$ is the effective sample size; suggesting that the bias is not converging to zero at the appropriate rate. The robust summaries tell a different story. The robust estimators of the SE (Med ASE, WMN ASE, and TMN ASE) were all in good agreement with the MAD. The ratio of median bias to median ASE (bias/SE) was within $1 / \sqrt{n_{\text{eff}}}$ for all estimators, indicating desirable convergent properties. The CP was improving as N increased, with the  CP* was close to the nominal 95\% for all estimators at N=17,000. 

While the coverage proportion was affected by non-convergent values, it was unaffected by outlying values. Removing datasets that resulted in non-convergence for any method (i.e., if any method did not converge for a given dataset, the dataset was removed when computing coverage proportion for all methods; denoted CP*) yielded increased coverage for all methods besides MICE.

\begin{table}[!h]
\centering
\caption{Estimating the cOR in a rare-outcome, high missing-data setting, n = 12000.\label{tab:synthetic_results_cOR_n12000}}
\centering
\fontsize{8.5}{10.5}\selectfont
\begin{threeparttable}
\begin{tabular}[t]{l>{\raggedleft\arraybackslash}p{0.3in}>{\raggedleft\arraybackslash}p{0.3in}>{\raggedleft\arraybackslash}p{0.3in}>{\raggedleft\arraybackslash}p{0.3in}>{\raggedleft\arraybackslash}p{0.22in}>{\raggedleft\arraybackslash}p{0.22in}>{\raggedleft\arraybackslash}p{0.22in}>{\raggedleft\arraybackslash}p{0.22in}>{\raggedleft\arraybackslash}p{0.22in}>{\raggedleft\arraybackslash}p{0.22in}>{\raggedleft\arraybackslash}p{0.22in}>{\raggedleft\arraybackslash}p{0.22in}>{\raggedleft\arraybackslash}p{0.22in}}
\toprule
Est & Mn bias & \textbf{Med bias} & WMn bias & TMn bias & ESE & \textbf{MAD} & Mean ASE & \textbf{Med ASE} & WMn ASE & TMn ASE & bias / SE & CP & \textbf{CP*}\\
\midrule
GR & -0.10 & -0.02 & -0.02 & -0.02 & 1.46 & 0.36 & 0.32 & 0.30 & 0.32 & 0.31 & 0.05 & 0.77 & 0.88\\
IPW & -0.23 & -0.10 & -0.10 & -0.09 & 2.22 & 0.91 & 0.80 & 0.77 & 0.80 & 0.79 & 0.12 & 0.90 & 0.91\\
MICE & 0.01 & 0.01 & 0.01 & 0.01 & 0.28 & 0.27 & 0.27 & 0.26 & 0.27 & 0.27 & 0.04 & 0.95 & 0.94\\
TMLE & -0.24 & -0.10 & -0.12 & -0.11 & 2.25 & 0.91 & 0.80 & 0.77 & 0.79 & 0.79 & 0.14 & 0.89 & 0.90\\
\bottomrule
\end{tabular}
\begin{tablenotes}
\item Abbreviations: mRD: marginal risk difference; Est: Estimator; Mn: Mean; Med: Median; WMn: Winsorized mean; TMn: trimmed mean; ESE: empirical standard error; MAD: median absolute deviation; ASE: asymptotic standard error; bias/SE: absolute Med bias / Med ASE; CP: coverage probability based on the ASE; CP*: coverage probability based on the ASE, where non-convergent datasets were dropped for all estimators; GR: generalized raking; IPW: inverse probability weighting; MICE: multiple imputation via chained equations; TMLE: targeted maximum likelihood estimation. Results are based on 2500 Monte-Carlo replications. 325 values were removed when computing GR performance due to non-convergence. Bolding indicates columns that we recommend using. The inverse square root of the effective sample size is 0.091.
\end{tablenotes}
\end{threeparttable}
\end{table}

\begin{table}[!h]
\centering
\caption{Estimating the cOR in a rare-outcome, high missing-data setting, n = 17000.\label{tab:synthetic_results_cOR_n17000}}
\centering
\fontsize{8.5}{10.5}\selectfont
\begin{threeparttable}
\begin{tabular}[t]{l>{\raggedleft\arraybackslash}p{0.3in}>{\raggedleft\arraybackslash}p{0.3in}>{\raggedleft\arraybackslash}p{0.3in}>{\raggedleft\arraybackslash}p{0.3in}>{\raggedleft\arraybackslash}p{0.22in}>{\raggedleft\arraybackslash}p{0.22in}>{\raggedleft\arraybackslash}p{0.22in}>{\raggedleft\arraybackslash}p{0.22in}>{\raggedleft\arraybackslash}p{0.22in}>{\raggedleft\arraybackslash}p{0.22in}>{\raggedleft\arraybackslash}p{0.22in}>{\raggedleft\arraybackslash}p{0.22in}>{\raggedleft\arraybackslash}p{0.22in}}
\toprule
Est & Mn bias & \textbf{Med bias} & WMn bias & TMn bias & ESE & \textbf{MAD} & Mean ASE & \textbf{Med ASE} & WMn ASE & TMn ASE & bias / SE & CP & \textbf{CP*}\\
\midrule
GR & -0.01 & -0.01 & -0.01 & -0.01 & 0.55 & 0.29 & 0.27 & 0.26 & 0.27 & 0.27 & 0.05 & 0.86 & 0.93\\
IPW & -0.06 & -0.05 & -0.05 & -0.05 & 0.99 & 0.73 & 0.67 & 0.65 & 0.67 & 0.66 & 0.07 & 0.92 & 0.93\\
MICE & 0.01 & 0.00 & 0.01 & 0.01 & 0.23 & 0.23 & 0.23 & 0.22 & 0.23 & 0.23 & 0.02 & 0.96 & 0.95\\
TMLE & -0.07 & -0.07 & -0.06 & -0.07 & 1.03 & 0.76 & 0.67 & 0.65 & 0.67 & 0.66 & 0.11 & 0.90 & 0.91\\
\bottomrule
\end{tabular}
\begin{tablenotes}
\item Abbreviations: mRR: marginal relative risk; Est: Estimator; Mn: Mean; Med: Median; WMn: Winsorized mean; TMn: trimmed mean; ESE: empirical standard error; MAD: median absolute deviation; ASE: asymptotic standard error; bias/SE: absolute Med bias / Med ASE; CP: coverage probability based on the ASE; CP*: coverage probability based on the ASE, where non-convergent datasets were dropped for all estimators; GR: generalized raking; IPW: inverse probability weighting; MICE: multiple imputation via chained equations; TMLE: targeted maximum likelihood estimation. Results are based on 2500 Monte-Carlo replications. 171 values were removed when computing GR performance due to non-convergence. Bolding indicates columns that we recommend using. The inverse square root of the effective sample size is 0.077.
\end{tablenotes}
\end{threeparttable}
\end{table}

\section{Convergence in distribution}

We argue that an asymptotic approximation is `good enough' if the probability of an importantly large discrepancy is sufficiently small.  This corresponds well to the Prohorov distance, which metrizes convergence in distribution. 

To construct the Prohorov distance between distributions $P$ and $Q$, define $A^\epsilon$ to be the set of points within $\epsilon$ of a set $A$, then write $d(P,Q)\leq\epsilon$ when for every open set $A$, we have $ P(A) < Q(A^\epsilon)+\epsilon$ and  $ Q(A) < P(A^\epsilon)+\epsilon$.  \citet{huber-book} defined `robust' estimators as those that are continuous in this metric (and more generally, continuous in the topology of convergence in distribution). 

The median and MAD, and the sample proportion estimator of confidence interval coverage, are continuous in the Prohorov metric and robust in Huber's sense, but  the sample moments are not.  Other robust estimators include trimmed means and standard deviations, and a wide collection of M--, L--, and R--estimators.

\section{Good data analysis practice}
The literature on robust statistics demonstrates clearly that small fractions of outliers can have a large impact on the distributions of moments. It is sometimes necessary to use estimators such as the mean that are sensitive to outliers, since robust location and scale estimators may not estimate the parameter of interest for the analysis. However, it is simpler to not use the mean for Monte Carlo distributions with the potential for extreme outliers unless there strong substantive reasons for caring about the mean rather than some other location summary. In many simulation studies, there is no particular substantive reason to prefer the mean over the median as a summary of the center of simulation results.  If the true mean is the quantity of interest, it is important to design a large enough simulation study to estimate it; as the logistic regression example shows, this can sometimes be challenging.

When an estimator performs badly in an unusual data set, the precise result is often sensitive to implementation details, making it still more difficult to summarise results appropriately and reproducibly. For example, the numerical results from  logistic regression software when $\hat\beta=\infty$ are sensitive to the convergence criterion: is convergence declared based on change in deviance or in likelihood score or in $\hat\beta$, and at what tolerance?

\section{Disadvantages to robust summaries}

There will be some settings where the finite-sample mean of an estimator is genuinely of interest; obviously the median is unsuitable in these settings.  As one example, a likelihood score has exactly zero expected value (but not median) even with finite sample size, and checking this by simulation is a useful validation technique \citep[eg][]{casexover-bias}. 

The median/MAD do have lower efficiency than the mean/standard deviation at the Normal, so they need a larger number of simulation replicates for the same precision if outliers are actually not a problem.  If computational resources are strongly constrained and a simulation sample size is chosen to achieve minimal acceptable precision, the mean/standard deviation could be preferred.  In our experience, however, simulation sample sizes are often not based on careful sample size calculations, and the mean/standard deviation can have arbitrarily low efficiency even in small (Prohorov) neighbourhoods of the Normal distribution.  

Finally, there is a wide range of possible robust estimators of location and scale, and some of these may be preferable to the median and MAD. On the other hand, it would be valuable to have a standard choice, to reduce researcher degrees-of-freedom \citep{researcher-df}. The median and MAD are computationally straightforward and well understood. They are also distinguished as the `most robust' $M$-estimators of location and scale by \citet[Section 2.4b]{hampel-book}, but it might be argued that summaries with a lower breakdown point and higher efficiency at the Normal would be better.

\section{Discussion}

We recommend median, MAD, and empirical confidence interval coverage for summarising simulation results unless there is a specific reason to choose something different. The median and MAD are commonly used and easy to interpret robust statistics that will perform similarly to the mean and SE when simulation estimates are well-behaved, with no outlier estimates. Focusing on the distributional quantities of the median and coverage probability will also provide more appropriate summaries of the performance characteristics of interest for the estimators being evaluated by the simulation, which relate to asymptotic convergence to the desired normal distribution.


\bibliographystyle{apalike}
\bibliography{interactnlmsample}
\end{document}